\documentclass[aip,reprint,amsmath,amssymb]{revtex4-1}

\usepackage{graphicx}
\usepackage{dcolumn}
\usepackage{bm}
\usepackage{amssymb}
\usepackage{natbib}

\begin{document}


\title{Evaluating the Applicability of the Fokker-Planck Equation in Polymer Translocation:
A Brownian Dynamics Study} 

\author{James M. Polson}
\email{jpolson@upei.ca}
\thanks{Corresponding author}
\author{Taylor R. Dunn}
\affiliation{%
Department of Physics, University of Prince Edward Island, 550 University Ave.,
Charlottetown, Prince Edward Island, C1A 4P3, Canada
}%

\date{\today}

\begin{abstract}
Brownian dynamics (BD) simulations are used to study the translocation dynamics of a coarse-grained 
polymer through a cylindrical nanopore. We consider the case of short polymers, with a polymer 
length, $N$, in the range $N=21$ -- 61. The rate of translocation 
is controlled by a tunable friction coefficient, $\gamma_{0{\rm p}}$, for monomers inside the nanopore. 
In the case of unforced translocation, the mean translocation time scales with polymer length as 
$\langle\tau_1\rangle\sim (N-N_{\rm p})^\alpha$,
where $N_{\rm p}$ is the average number of monomers in the nanopore. The exponent approaches the 
value $\alpha=2$ when the pore friction is sufficiently high, in accord with the prediction for
the case of the quasi-static regime where pore friction dominates. In the case of forced
translocation, the polymer chain is stretched and compressed on the {\it cis} and {\it trans}
sides, respectively, for low $\gamma_{0{\rm p}}$. However, the chain approaches conformational
quasi-equilibrium for sufficiently large $\gamma_{0{\rm p}}$. In this limit the observed scaling of
$\langle\tau_1\rangle$ with driving force and chain length supports the FP prediction
that $\langle\tau\rangle\propto N/f_{\rm d}$ for sufficiently strong driving force.
Monte Carlo simulations are used to calculate translocation free energy functions for the
system. The free energies are used with the Fokker-Planck equation to calculate translocation
time distributions. At sufficiently high $\gamma_{0{\rm p}}$, the predicted distributions are 
in excellent agreement with those calculated from the BD simulations. Thus, the FP equation 
provides a valid description of translocation dynamics for sufficiently high pore friction 
for the range of polymer lengths considered here. Increasing $N$ will require a corresponding 
increase in pore friction to maintain the validity of the FP approach. Outside the regime
of low $N$ and high pore friction, the polymer is out of equilibrium, and the FP approach
is not valid.
\end{abstract}

\maketitle

\section{Introduction}
\label{sec:intro}

The translocation of polymers through nanometre-sized pores and channels is an important 
and complex process that has been the subject of a large body of research in 
recent years.\cite{Muthukumar_book,Panja_2013,Milchev_2011} In part, this interest
is due to its relevance to fundamental biological processes, including DNA
and RNA transport across nuclear pores, protein transport through membrane channels,
genome packing in bacteriophages, and so on.\cite{Alberts_book,Lodish_book}  In addition, 
the development of experimental techniques to detect and characterize translocation at 
the single-molecule level has been a major force driving activity in this field.
Since the pioneering study of Kasianowicz {\it et al}.\cite{Kasianowicz_1996}
there has been substantial effort devoted to the refinement of these methods using 
biological\cite{Akeson_1999,Meller_2000,Meller_2003,Wang_2004,Butler_2006,Wong_2008,Wong_2010}
and solid-state\cite{Li_2003,Chen_2004, Fologea_2005a,Fologea_2005b,Storm_2005,Storm_2005b,Purnell_2008}
nanopores. Much of this work is inspired by its potential technological applications, notably
the development of a nanopore-based device for rapid DNA sequencing.%
\cite{Branton_2008,Venkatesan_2011,Wanunu_2012,Yang_2013} 

Numerous theoretical studies have investigated different aspects of polymer translocation
using a variety of analytical and computer simulation methods. One of the first 
theoretical approaches to translocation was developed by Sung and Park,\cite{Sung_1996} 
and Muthukumar.\cite{Muthukumar_1999} Here, translocation is treated as a process in
which the polymer maintains conformational quasi-equilibrium throughout, and the dynamics
are reduced to monitoring the time-dependence of the translocation coordinate, $s$, the
position of the monomer inside the pore. The method involves solving the Fokker-Planck (FP) 
equation using an analytical approximation for the equilibrium free energy function $F(s)$. 
The method has been reviewed in detail in Ref.~\onlinecite{Muthukumar_book}.
Numerous studies have employed the FP approach to study translocation.\cite{%
Sung_1996,Muthukumar_1999,Slonkina_2003,Muthukumar_2003,%
Slonkina_2003,Kong_2004,Kejian_2006,Gauthier_2008a,Gauthier_2008b,Wong_2008,Mohan_2010,deHaan_2011a,%
Yang_2012,Zhang_2013} One notable prediction is form of the scaling of
the mean first passage time, $\langle\tau\rangle$, with the polymer length, $N$.
Assuming that pore friction dominates, it follows that $\langle\tau\rangle\sim N^2$ 
for unforced translocation,\cite{Muthukumar_1999,deHaan_2011a} 
and $\langle\tau\rangle\sim N/f_{\rm d}$ in the presence of a strong driving force, $f_{\rm d}$.
The latter result is consistent with experimental results for ss-DNA translocation
through $\alpha$-hemolysin pores,\cite{Muthukumar_book,Kasianowicz_1996} but is inconsistent with some
results for DNA translocation through solid-state nanopores.\cite{Storm_2005,Wanunu_2008}

The quasi-equilibrium condition, if it holds at all, must eventually break down for sufficiently 
long polymers.\cite{Chuang_2001,Kantor_2004} Non-equilibrium conformational behaviour has
been been directly observed in computer simulations.\cite{Luo_2006b,Gauthier_2009,Bhattacharya_2009,%
Luo_2010,Bhattacharya_2010,Dubbeldam_2012,Feng_2012} In addition, most simulation studies have 
yielded translocation time scaling exponent values inconsistent with those predicted by the FP method for
the quasi-static regime.\cite{Chuang_2001,Milchev_2004,Kantor_2004,Tsuchiya_2007,Luo_2006a,Wolterink_2006,%
Guillouzic_2006,Dubbeldam_2007a,Dubbeldam_2007b,Huopaniemi_2007,Panja_2007,Panja_2008a,%
Wei_2007,Luo_2008a,Vocks_2008,Lehtola_2008,Lehtola_2009,Luo_2009,%
Gauthier_2009,Bhattacharya_2009,Kapahnke_2010,deHaan_2010,Dubbeldam_2011,deHaan_2012b,%
Ikonen_2012a,Edmonds_2013} Conformational quasi-equilibrium is only possible when
the polymer relaxation time is sufficiently short compared to the translocation time, and
this condition is apparently not satisfied in most simulation studies.
A notable exception was reported recently in the Langevin dynamics study of non-driven 
translocation by de~Haan and Slater.\cite{deHaan_2012b} Here, the conformational relaxation 
time was reduced by decreasing the the solvent viscosity. At sufficiently low
viscosity and fast chain relaxation, a scaling exponent of $\alpha=2$ was measured for
polymer chains with lengths in the range $N=25$--99.  This was the first confirmation using
simulation methods that the FP approach is valid in the limit of sufficiently fast 
subchain relaxation. The other means of maintaining quasi-equilibrium is by increasing
the pore friction and thus slowing the translocation rate. This was demonstrated
in a recent study by one of us, in which Monte Carlo (MC) dynamics simulations were
used to study the translocation of a hard-sphere chain through a cylindrical nanopore.\cite{Polson_2013b}
In that study, the translocation rate was decreased by decreasing the nanopore width,
which increases the effective friction of monomers in the nanopore. The translocation
time distributions were compared with those calculated by solving the FP equation
using explicitly calculated free energy functions.\cite{Polson_2013} 
The theoretical distributions were fit to the simulation distributions by
adjusting the effective diffusion coefficient $D$ appearing in the FP equation.
In the limit of sufficiently high pore friction, the quality of the theoretical predictions
was generally excellent for physically meaningful values of $D$.

Generally, the quasi-equilibrium regime is determined by the pore friction and (for driven
translocation) the driving force, which affect the translocation rate, as well as polymer
length and solvent viscosity, which affect the relaxation rate.
Outside this regime, various other theoretical models have been developed to describe the
dynamics in the out-of-equilibrium case for non-driven\cite{Panja_2007,Dubbeldam_2007a,deHaan_2012b} 
and driven\cite{Dubbeldam_2007b,Vocks_2008,Sakaue_2007,Sakaue_2010,Saito_2011,Rowghanian_2011,%
Dubbeldam_2012,Ikonen_2012a,Ikonen_2012b,Ikonen_2012c} translocation. 
Especially noteworthy is the tension propagation (TP) theory for driven translocation pioneered by 
Sakaue\cite{Sakaue_2007,Sakaue_2010,Saito_2011,Saito_2012a} and further refined
by others.\cite{Rowghanian_2011,Dubbeldam_2012} 
%
%
This model was recently extended by Ikonen {\it et al.}.\cite{Ikonen_2012a,Ikonen_2012b,%
Ikonen_2012c,Ikonen_2013} Unlike the original theory, their Brownian dynamics tension 
propagation (BDTP) approach is valid for chains of finite length, and also includes
a contribution to the total force due to pore friction. Using this method, it was
demonstrated that the scaling exponent $\alpha$ exhibits a pronounced dependence
on chain length and approached the asymptotic limit of $\alpha=1+\nu\approx 1.6$ only at very
large $N$, thus showing that the wide range of $\alpha$ measured in various simulation
studies is a finite-size effect.\cite{Ikonen_2012a,Ikonen_2012b} 
In addition, $\alpha$ was found to decrease with decreasing $N$ and increasing 
pore friction, apparently tending toward the value of $\alpha=1$, which is the FP 
prediction in the case of sufficiently strong driving force.\cite{Ikonen_2012b}
Thus, the FP predictions may be a limiting case of the more general BDTP theory.

In the present study, we use Brownian dynamics simulation to study translocation of a 
finitely extensible nonlinear elastic (FENE) model polymer\cite{Bird_1987} through 
a cylindrical nanopore. The translocation rate is controlled by explicit adjustment 
of the friction of monomers inside the nanopore. This work is largely a 
continuation of the work in Ref.~\onlinecite{Polson_2013b}. In addition, it is
closely related to the theoretical study of Ref.~\onlinecite{Ikonen_2012b},
which also studied the effects of pore friction.  We examine both unforced and forced 
translocation and analyze the translocation time distributions using the FP method together with explicitly
calculated free energy functions. As expected, quasi-static conditions are observed for 
sufficiently high pore friction and slow translocation, in which limit, however, 
the simulations became computationally expensive. Consequently, we restrict this
study to relatively short polymer chains.  In this limit, we find that the translocation 
dynamics are consistent with FP predictions for sufficiently high nanopore friction.

The remainder of this article is organized as follows. 
In Sec.~\ref{sec:model} we describe the model used in the simulations.
In Sec~\ref{sec:theory}, we review the essential equations of the FP
formalism. We also establish the connection between the diffusion coefficient
appearing in the FP equation and the relevant parameters of the molecular
model. In Sec.~\ref{sec:methods} we describe the MC simulations used to 
calculate the translocation free energy functions, the details of the
BD simulations, and the calculations using the FP equation.
In Sec.~\ref{sec:results} the results of the simulations and theoretical
calculations are presented. The significance of those results
is discussed in Sec.~\ref{sec:conclusions}.

\section{Model}
\label{sec:model}

The polymer is modeled as a flexible chain of $N$ spherical particles.
All particles interact with the repulsive Lennard-Jones potential:
\begin{equation}
\label{eq:upair}
u(r) =
\begin{cases}
u_{\rm LJ}(r) - u_{\rm LJ}(r_{\rm min}), ~~~~ r < r_{\rm min} \\
0, ~~~~~~~~~~~~~~~~~~~~~~~~~~~ r \geq r_{\rm min}
\end{cases}
\end{equation}
where
\begin{eqnarray}
u_{\rm LJ}(r) = 4\epsilon \left[\left(\sigma/r\right)^{12}-\left(\sigma/r\right)^6\right],
\end{eqnarray}
and where $r_{\rm min}= 2^{1/6}\sigma$.
Bonded monomers also interact with the FENE potential,
\begin{eqnarray}
u(r) = -{\textstyle\frac{1}{2}} k r_0^2 \ln\left(1-(r/r_0)^2\right),
\end{eqnarray}
where $r$ is the interparticle distance, and where $k\sigma^2/\epsilon=30$
and $r_0=1.5\sigma$.

The nanopore is a cylindrical hole of radius $R$ in a barrier of length $L$.
We choose the center of the cylinder to be along the $z$ axis, with 
$z=0$ at one end of the pore and $z=L$ at the other. Monomers interact with 
an interaction site at the nearest point on the barrier or nanopore wall
with the potential of Eq.~(\ref{eq:upair}). Note that the pore radius is 
defined such that $R+0.5\sigma$ is the distance between the nearest interaction 
site on the pore wall to the centre of the pore.  Monomers inside the pore are subject 
to a constant driving force of $f_{\rm d}$ for $z \in [0,L]$.
Thus, the associated potential energy is:
\begin{eqnarray}
u_{\rm d} = 
\begin{cases}
0, ~~ z < 0 \\
f_{\rm d} z, ~~ 0 \leq z \leq L \\
f_{\rm d} L, ~~ z \geq L
\end{cases}
\label{eq:ud}
\end{eqnarray}

\begin{figure}[!ht]
\includegraphics[width=0.35\textwidth]{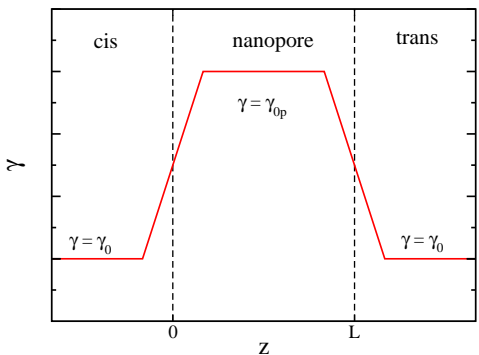}
\caption{Variation of the monomer friction coefficient, $\gamma$, with $z$.
Note that $\gamma$ varies continuously from $\gamma_0$ outside the pore
to $\gamma_{0{\rm p}}$ inside the pore. The transition zones are centered
at the pore edges and have a width of $\Delta z = \sigma$.
}
\label{fig:gamma_pic}
\end{figure}

For the Brownian dynamics (BD) algorithm employed in this study and described below in 
Section~\ref{subsec:theory-dynamics},
each particle has an associated friction coefficient, $\gamma$.  The value of $\gamma$ 
depends on the position of the monomer in the pore. Outside the pore, the coefficient
has a value $\gamma_0$, while inside the pore, its value is $\gamma_{0{\rm p}}$. The
value of $\gamma_{0{\rm p}}$ controls the rate of translocation.
In this study, we consider the range where $\gamma_{0{\rm p}} \geq \gamma_0$, i.e.
the friction in the pore exceeds that for a monomer outside the pore.
To avoid problems with the BD algorithm, $\gamma$ is chosen to be a continuous
function of $z$. The function has the form:
\begin{eqnarray}
\gamma =
\begin{cases}
\gamma_0, \hfill z \leq -\frac{\sigma}{2} \\
\gamma_0 + m(z+\frac{\sigma}{2}), \hfill -\frac{\sigma}{2}\leq z \leq \frac{\sigma}{2} \\
\gamma_{0{\rm p}}, \hfill  \frac{\sigma}{2} \leq z \leq L-\frac{\sigma}{2}\\
\gamma_0 + m(L+\frac{\sigma}{2}-z), ~~ \hfill L-\frac{\sigma}{2}\leq z \leq L+\frac{\sigma}{2} \\
\gamma_0, \hfill z \geq L + \frac{\sigma}{2}
\end{cases}
\label{eq:gamma_def}
\end{eqnarray}
where $m\equiv (\gamma_{0{\rm p}}-\gamma_0)/\sigma$.
The variation of $\gamma$ with $z$ is illustrated in Fig.~\ref{fig:gamma_pic}.

\begin{figure}[!ht]
\includegraphics[width=0.35\textwidth]{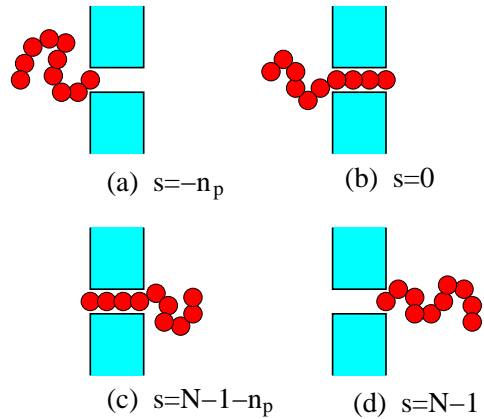}
\caption{Illustration showing translocation coordinate $s$
at different points during translocation. The pore-filling stage lies
between states (a) and (b), the transfer (i.e. filled pore) stage lies
between states (b) and (c), and the pore emptying stage lies between
states (c) and (d). $n_{\rm p}$ is the average number of bonds spanning
the nanopore, and $N$ is the number of monomers.
}
\label{fig:spic}
\end{figure}

The degree to which the polymer has translocated across the nanopore is quantified
using a translocation coordinate, $s$. This coordinate is defined in a manner similar to
that of most other translocation studies, though with some slight differences.
It is defined in relation to another coordinate, $s'$, which is chosen to be
the number of bonds that have crossed the mid-point of the nanopore at $z=L/2$. Typically,
one bond crosses this point, and this bond contributes to $s'$ the fraction
that lies on the {\it trans} side of the point. This is determined by the $z$ coordinates
of the monomers connected by this bond.  Note that $s'$ is a continuous variable in
the range $s'\in[0,N-1]$. The coordinate $s$ is given by $s=s'-n_{\rm p}/2$,
where $n_{\rm p}$ is the number of bonds spanning the nanopore when filled.
It is defined $n_{\rm p}\equiv L/\Delta z_{\rm p}$,
where $\Delta z_{\rm p}$ is average bond length projected along the nanopore $z$ axis.
In this study, the pore is sufficiently narrow that the polymer is strongly aligned
inside, and simulations yield a value of $\Delta z_{\rm p}=0.96\sigma$, which
is also the equilibrium bond length of the polymer.
From this definition of $s$, the range $[0,N-1]$ corresponds to the translocation stage
when the nanopore is filled. To complete the definition of $s$ for the case
where the pore is less than half full (i.e. no bonds cross the mid-point of the pore), 
we define $s=-(n_{\rm p}/2-z_1/\Delta z_{\rm p})$, where $z_1$ is the coordinate of the end monomer.
Likewise, in the case where less than half filled pore during emptying, $s=N-1-(z_N-L)/\Delta z_{\rm p}$,
where $z_N$ is the coordinate of the other end. With this definition of $s$, the following 
domains can be identified: (i) the pore filling stage for $s\in[-n_{\rm p},0]$;
(ii) the filled pore stage for $s\in[0,N-1-n_{\rm p}]$; and the pore-emptying stage
for $s\in[N-1-n_{\rm p},N-1]$. This is illustrated in Fig.~\ref{fig:spic}.

In our previous work,\cite{Polson_2013,Polson_2013b} we employed a different 
translocation coordinate, $Q$, which was defined solely in terms of the coordinates of 
the monomers. Thus, unlike the coordinate $s$, its calculation is straightforward
and does not require {\it a posteriori} knowledge of the quantity $\Delta z_{\rm p}$. However, it 
has the disadvantage that it does not vary linearly with monomer displacement along the 
pore during the pore filling and emptying stages. By contrast, a given change in $s$ 
corresponds to the same displacement for monomers inside the pore for {\it all} stages
of translocation.

\section{Fokker-Planck Equation and the Friction Coefficient}
\label{sec:theory}

In the quasi-static limit, the time-dependent translocation probability distribution,
${\cal P}(s,t)$, is governed by the Fokker-Planck equation.
The general form of the equation is\cite{Muthukumar_book}
\begin{eqnarray}
\frac{\partial{\cal P}(s,t)}{\partial t} & = &
\frac{\partial}{\partial s} \left(\frac{D}{k_{\rm B}T} \frac{\partial F}{\partial s}
{\cal P}(s,t)\right) + \frac{\partial}{\partial s} D \frac{\partial {\cal P}(s,t)}{\partial s},
\label{eq:FP}
\end{eqnarray}
where $k_{\rm B}$ is Boltzmann's constant, $T$ is temperature, and
$F(s)$ is the equilibrium translocation free energy function. 
In this study, $F(s)$ is calculated using the MC simulation
method described in Ref.~\onlinecite{Polson_2013}
and in Section~\ref{subsec:theory-free-energy} below.
In addition, $D$ is an effective diffusion coefficient that can be used to define 
the polymer friction parameter $\gamma_{\rm pol}$ using the Einstein relation:
\begin{eqnarray}
D = \frac{k_{\rm B}T}{\gamma_{\rm pol}}.
\end{eqnarray}
Generally, $D$ and $\gamma_{\rm pol}$ depend on the translocation position, $s$.

To establish a relationship between $\gamma_{\rm pol} $ and the simulation parameter 
$\gamma$ defined in Eq.~(\ref{eq:gamma_def}), we consider only the case of very strong
pore friction. In this limit, the overall friction of the polymer is dominated
by the friction of the monomers inside the pore. If $N_{\rm eff}$ monomers lie inside
the pore, then the total friction of these monomers (and thus the polymer)
is approximately 
\begin{eqnarray}
\gamma_{\rm pol} = N_{\rm eff} \gamma_{0{\rm p}}, 
\label{eq:gamma_z}
\end{eqnarray}
where $\gamma_{0{\rm p}}$ was defined earlier. The quantity $N_{\rm eff}$ can be estimated
\begin{eqnarray}
N_{\rm eff} \approx L/\Delta z_{\rm p},
\label{eq:NeffLas}
\end{eqnarray}
where $\Delta z_{\rm p}$ is the average distance along $z$ between bonded monomers
inside the pore. For the nanopore dimensions used here ($L=3\sigma$ and $R=0.65\sigma$), 
$\Delta z_{\rm p}=0.96\sigma$ for monomers pairs located with $z\in[0,L]$,
for which $N_{\rm eff} = 3.125$. However, this turns out to be a slight underestimate 
of the true pore friction $\gamma_z$. A more accurate value can be determined by 
calculating the total friction of the pore monomers using the expression for the 
monomer friction in Eq.~(\ref{eq:gamma_def}), and averaging over coordinates of the
monomers for different configurations of the system. These configurations can
be obtained from equilibrium MC simulations described below in 
Section~\ref{subsec:theory-free-energy}. For a pore of dimensions $L=3\sigma$ 
and $R=0.65\sigma$, we find $N_{\rm eff}\approx 3.2$.

In practice, the value of $N_{\rm eff}$ is defined by Eq.~(\ref{eq:gamma_z}), where 
$\gamma_{\rm pol}$ is the polymer friction coefficient that provides the best fit to 
the simulation data. In this sense, $N_{\rm eff}$ is the effective number of monomers 
that are subject to the friction in the nanopore. Significant deviations from the estimate
in Eq.~(\ref{eq:NeffLas}) can be interpreted as a breakdown in the underlying
assumptions of conformational equilibrium and/or negligible friction of
monomers outside the pore. Furthermore, in the strong pore friction regime
where $\gamma_{0{\rm p}}$ is sufficiently large, $\gamma_{\rm pol}$ and, thus, $N_{\rm eff}$
should be independent of $\gamma_{0{\rm p}}$. We use this fact to locate the
quasi-equilibrium regime.

Consider the case of a partially filled nanopore, again in the limit of strong pore friction,
$\gamma_{0\rm p} \gg \gamma_0$ and a strong polymer alignment in the pore. 
Here, $\gamma_{\rm pol}$ is expected to vary with $s$. For example, $\gamma_{\rm pol}$ should
decrease with decreasing $s$ as the pore empties monomers on the {\it cis} side.
However, from the variation of $\gamma$ with position in Eq.~(\ref{eq:gamma_def}) and
illustrated in Fig.~\ref{fig:gamma_pic} it can be shown that the filled-pore value
of $\gamma_{\rm pol}$ extends approximately to $s=-0.5\sigma/\Delta z_{\rm p}$ on the {\it cis} 
side and decreases approximately linearly with decreasing $s$ to a value of 
$\gamma_{\rm pol}=(\gamma_{0\rm p}+\gamma_0)/2$ at $s=-n_{\rm p}$. The value on 
the {\it trans} side is related by symmetry.  MC simulations used to calculate 
$\gamma_{\rm pol}(s)$ using Eq.~(\ref{eq:gamma_def}) and averaging over polymer 
configurations show very good agreement with this approximation.
We use this approximation for FP calculations that consider the pore-emptying
stages.

Under the assumption that the polymer maintains a state of conformational 
quasi-equilibrium during translocation and that the pore friction dominates,
the relations derived above can be used to solve the FP equation. One important
quantity of interest that emerges from such a calculation is the translocation
first passage time. Consider a domain bounded by $s=a$ and $s=b$.
The first passage time, $\tau$, for a polymer located at $s_0 \in [a,b]$
at time $t=0$ to reach either boundary has a probability distribution given 
by\cite{Muthukumar_book}
\begin{eqnarray}
g(\tau;s_0) = -\frac{d}{d\tau} \int_{a}^{b} ds \,p(s,\tau;s_0,0),
\label{eq:gQ0tau}
\end{eqnarray}
where $p(s,t;s',t')$ is the conditional probability that the polymer reaches
a value $s$ at time $t$ given that it had a value of $s'$ at an earlier time $t'$.
For a Markov process, $p(s,t;s',t')$ also satisfies Eq.~(\ref{eq:FP}). For most calculations 
in this study, we choose $a=0$ and $b=N-1-n_{\rm p}$, i.e the points at which monomers first empty 
the {\it trans} and {\it cis} sides, as illustrated in Fig.~\ref{fig:spic}(b) and (c).
Thus, $[a,b]$ is the maximum range of $s$ for a filled nanopore.

\section{Methods}
\label{sec:methods}

\subsection{Free Energy Calculations}
\label{subsec:theory-free-energy}

Monte Carlo simulations employing the Metropolis algorithm and the self-consistent 
multiple histogram (SCMH) method\cite{Frenkel} were used to calculate the free 
energy functions for the polymer-nanopore model described in Section~\ref{sec:model}. 
A detailed description of the methodology is given in our recent article
on the translocation of hard-sphere chains.\cite{Polson_2013} The MC calculations
in this work are similar to those of Ref.~\onlinecite{Polson_2013}. However, 
the choice of definition of the translocation coordinate here requires that the calculations 
be performed in two stages. In the first stage, we evaluate the free energy
function in the range $s\in[-n_{\rm p}/2,N-1-n_{\rm p}/2]$. The bounds of this range
correspond to the cases where an end monomer located at the centre of the pore. 
Thus, it includes the filled-pore stage, as well as half the filling and emptying
stages. In this range, $s'=s-n_{\rm p}$ is the number of bonds that have crossed
the midpoint of the pore, as described in Section~\ref{sec:model}. In the second stage,
we calculate the free energy in the filling and emptying stages using the definition
of $s$ in terms of the position of the end monomers. The different pieces of the free energy
functions are fitted together by imposing continuity at $s=-n_{\rm p}/2$ and $s=N-1-n_{\rm p}/2$.

To implement the SCMH method, we employ a ``window potential''
\begin{eqnarray}
{W_i(s)}=\begin{cases} \infty, \hspace{8mm} s<s_i^{\rm min} \cr 0,
\hspace{1cm} s_i^{\rm min}<s<s_i^{\rm max} \cr \infty, \hspace{8mm} s>s_i^{\rm max} \cr
\end{cases}
\label{eq:winPot}
\end{eqnarray}
where $s_i^{\rm min}$ and $s_i^{\rm max}$ are the limits that define the range of $s$
for the $i$-th window. These restrictions on $s$ are part of the acceptance criteria
for the MC trial moves,
and any proposed move which produces a configuration with a $s$ value outside this range
is rejected. Within each window of $s$, a probability distribution $p_i(s)$ is
calculated in the simulation. The window potential width,
$\Delta s \equiv s_i^{\rm max} - s_i^{\rm min}$, is chosen to be sufficiently small
that the variation in $F$ does not exceed a few $k_{\rm B}T$. Adjacent windows overlap,
and the SCMH algorithm uses the $p_i(s)$ histograms to reconstruct the unbiased distribution,
$P(s)$. For further details, see Ref.~\onlinecite{Polson_2013}.

The windows are chosen to overlap with half of the adjacent window,
such that $s^{\rm max}_{i} = s^{\rm min}_{i+2}$. In the first stage of the calculation,
the window widths were typically $\Delta s/(N-1) = 0.02$, which required a total of 
$n_{\rm win}=99$ overlapping windows to span the range $s\in [-n_{\rm p}/2,N-1-n_{\rm p}/2]$. 
In addition, individual probability histograms were constructed using the binning technique. We 
used 20 bins per histogram and so the sampling bin width was $\delta s/(N-1)=0.001$. For calculation
of the free energy for values of $s$ outside this range, the same parameters were used, except
that a smaller number of windows was required. 

The polymer configurations were generated by a random choice of three different particle 
moves.  In one case,  randomly chosen monomers were randomly displaced in each dimension.
The displacement was chosen from a uniform distribution
in the range $[-\Delta_{\rm max},\Delta_{\rm max}]$. The trial moves were accepted with a
probability $p_{\rm acc}=\min(1,e^{-\Delta E/k_{\rm B}T})$, where $\Delta E$ is
the energy difference between the trial and current states. The maximum displacement 
parameter was chosen to be $\Delta_{\rm max}=0.07\sigma$, which yielded an acceptance 
probability of approximately 46\%. In addition to single-monomer displacement, we also
used crankshaft rotations, as well as uniform random translation of all monomers collectively.
The maximum angular and translational displacements were chosen to yield comparably
reasonable acceptance ratios. At the beginning of each simulation, the polymer was placed 
in a linear conformation at a position in the pore such that $s\in [s_i^{\rm min},s_i^{\rm max}]$ 
for window $i$. The polymer was
equilibrated for an appropriate period prior to the sampling for the calculation of 
the histograms.  As an illustration, for a chain of length $N=41$, the equilibration time
was $2\times 10^7$ MC cycles, and the production run time was $8\times 10^8$ MC cycles.
A MC cycle corresponds to an attempt to move each monomer an average of one time for
each type of move.

\subsection{Brownian Dynamics Simulations}
\label{subsec:theory-dynamics}

We employ the BD simulation method to calculate translocation time distributions.
For this method, the coordinates of the {\it i}th particle are advanced 
through a time $\Delta t$ according to the algorithm:
\begin{eqnarray}
x_i(t+\Delta t) & = & x_i(t) 
+ \left( \frac{f_{i,x}}{\gamma_i} + k_{\rm B}T\frac{d \gamma_i^{-1}}{dz}\right) \Delta t 
\nonumber\\
&&+ \sqrt{2 k_{\rm B}T \Delta t/\gamma_i} \Delta w,
\label{eq:BDeq}
\end{eqnarray}
and likewise for $y_i$ and $z_i$.
Here, $f_{i,x}$ is the $x$-component of the conservative force on particle $i$.
In addition, $\Delta w$ is a random quantity drawn from a Gaussian of unit variance.
The term proportional to the derivative $d\gamma_i^{-1}/dz$ accounts for the
variation of the friction during the passage of a monomer through the pore. In
practice, it is non-zero only near the {\it cis} and {\it trans} edges of
the pore, as is evident in Fig.~\ref{fig:gamma_pic} and Eq.~(\ref{eq:gamma_def}).

For each translocation event simulation, the polymer was initially placed in a linear
conformation at a position corresponding to some desired initial coordinate, $s_0$.
The monomer closest to the centre of the nanopore was tethered to its
initial position with a strong harmonic ``pinning'' force, and the coordinates of 
all the monomers evolved according to Eq.~(\ref{eq:BDeq}) until conformational
equilibrium was reached.  The time required for equilibration was dependent on the 
chain length, $N$, and on $s_0$. Generally, longer {\it cis} or {\it trans} subchains 
required longer equilibration times.  An appropriate time was determined by measuring 
the conformational correlation times measured for chains tethered to a
flat surface in a series of separate simulations. As expected, the correlation
times for the tethered chains scaled as $N^{1+2\nu}\approx N^{2.2}$, the same
as that for a free Rouse chain.  The equilibration time for the 
translocation simulations was chosen to be several times longer than the 
correlation time for a tethered chain of a length equal to that of 
the longer of the two subchains.

Following equilibration, the pinning force was turned off, and the dynamics
of the polymer proceeded until the first time at which all monomers exited
the pore on either the {\it cis} or {\it trans} side. We call the time taken
for this complete emptying of the nanopore the translocation time, $\tau$.
In addition to this, we measure the first time, $\tau_1$, when all monomers 
have completely emptied either the {\it cis} or {\it trans} compartment.
Clearly, $\tau_1 < \tau$. In this study, we focus mainly on the time $\tau_1$.

For any given set of system parameter values, many different translocation simulations
were carried out. The number of translocation events was in the range of $10^3$--$10^5$.
These simulations were used to calculate mean translocation time, $\langle\tau_1\rangle$, 
and translocation time distributions, $g(\tau_1)$.

All simulations were carried out at a temperature $k_{\rm B}T/\epsilon=1$.
The time step used in Eq.~(\ref{eq:BDeq}) was $\Delta t = 0.0001 \gamma_0\sigma^2/\epsilon$.
Most calculations used a nanopore of dimensions $L=3\sigma$ and $R=0.65\sigma$.

\subsection{Theoretical Calculations}
\label{subsec:theory-calculations}

Using the free energy function calculated from the MC simulations, 
Eq.~(\ref{eq:FP}) was solved for a selected initial coordinate $s_0$. This calculation
was carried out using a standard finite-difference method with a ``spatial'' increment 
of $\Delta s/(N-1)=0.001$ and a time increment of $\Delta t=0.0001\gamma_0\sigma^2/\epsilon$. 
Adsorbing boundary conditions at $s=0$ and $s=N-1-n_{\rm p}$ were used in calculations
for $\tau_1$. The calculation yielded a histogram approximation for $P(s,t)$, or, equivalently, 
the quantity $p(s,t;s_0,0)$ in Eq.~(\ref{eq:gQ0tau}).  The distribution of first passage
times, $g(\tau_1;s_0)$, was calculated by solving Eq.~(\ref{eq:gQ0tau}). The integral
was solved numerically using the trapezoid rule, and the derivative was calculated
using a finite-difference approximation. The distribution function is dependent
on the parameter $N_{\rm eff}$. This quantity was adjusted to provide the best
fit to the distribution measured directly from the BD simulations.

\subsection{Units}
\label{subsec:units}

For the model used in this study, the defining quantities of measure are $\sigma$, 
$\epsilon$ and $\gamma_0$.  Thus, length is measured in units of $\sigma$, 
energy in units of $\epsilon$, friction in units of $\gamma_0$, force in units of 
$\epsilon/\sigma$, and time in units of $\gamma_0\sigma^2/\epsilon$.
The results below are presented using this set of units.

\section{Results}
\label{sec:results}

\begin{figure}[!ht]
\includegraphics[width=0.40\textwidth]{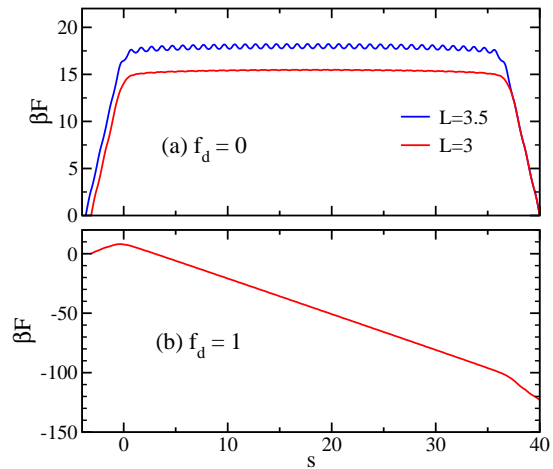}
\caption{Free energy functions for a polymer of length $N=41$ and a pore of 
radius $R=0.65$. Results are shown in (a) for non-driven translocation (i.e. $f_{\rm d}=0$) 
for pore lengths of $L=3.5$ and $L=3$. The results in (b) are for driven 
translocation for a driving force of $f_{\rm d}=1$ and a pore length of $L=3$.}
\label{fig:F.N41.L3.R0.65}
\end{figure}

Free energy functions were calculated using the MC method described in 
Section~\ref{subsec:theory-free-energy}. 
Sample free energy functions are shown in Fig.~\ref{fig:F.N41.L3.R0.65} 
for a polymer of length $N=41$ in a nanopore of radius $R=0.65$. 
The curves in Fig.~\ref{fig:F.N41.L3.R0.65}(a) correspond to the case of 
unforced translocation. Data for pore lengths of $L=3$ and $L=3.5$ are shown. 
The curves are qualitatively similar to those calculated for translocation
of hard-sphere chains in Ref.~\onlinecite{Polson_2013}. The sharp increase
and decrease of the function on either side correspond (mostly) to the pore-filling
and pore-emptying stages of translocation, respectively. The wide and slightly curved
plateau corresponds to the transfer stage, where the lengths of the {\it cis}
and {\it trans} subchains decrease and increase, respectively, as monomers move
through a filled pore. The symmetry of the functions reflects
the underlying symmetry of the model in the absence of a driving force.
The free energy barrier is higher for $L=3.5$ than for $L=3$ since a longer
nanopore corresponds to greater confinement, and thus lower conformational
entropy, in the filled-pore state. The ratio of the barrier heights is
$\approx 0.86$, which is equal to the ratio of the nanopore lengths. 
This is in accord with the expectation that the free energy barrier height 
be proportional to the pore length.\cite{Polson_2013} In addition, the $L=3.5$ 
curve has oscillations with a period of $\Delta s=1$ and an amplitude of 
$\beta\Delta F_{\rm osc}\approx 0.8$.  This feature is also present in the 
$L=3$ curve, but the amplitude is much smaller and the oscillations are barely detectable 
in the plot.  These oscillations are an artifact of the model due to the variation of the 
orientational entropy of bonds passing through the {\it cis} and {\it trans} edges of the 
pore.  This effect was discussed at length in Ref.~\onlinecite{Polson_2013}.
The variation of $\Delta F_{\rm osc}$ with $L$ follows a similar trend
as in that study. In addition, the oscillation amplitude decreases with
increasing $R$, in agreement with the results of Ref.~\onlinecite{Polson_2013}
(data not shown). Figure~\ref{fig:F.N41.L3.R0.65}(b) shows a free energy curve for driven
translocation for a driving force of $f_{\rm d}=1$. This force causes the
{\it trans} side to be energetically favoured to a considerable degree
relative to the {\it cis} side.

Let us consider further the free energy functions for unforced translocation.
The derivation of the standard analytical approximation for the entropic
barrier employs the result that the partition function for a self-avoiding
polymer tethered to a hard wall has the form\cite{deGennes}
\begin{eqnarray}
Z = \left(e^{-\beta\mu}\right)^n n^{\lambda-1},
\label{eq:Z}
\end{eqnarray}
where $n$ is the length of the chain and $\mu$ is the chemical potential per monomer.
In addition, the scaling exponent $\lambda$  has a value of $\lambda\approx 0.69$ for
a self-avoiding flexible polymer.\cite{Eisenriegler_1993} For a polymer translocating 
through a short nanopore, the configurational partition function is a product of two 
terms, each of the form of Eq.~(\ref{eq:Z}), for the {\it cis} and {\it trans} 
subchains.\cite{Muthukumar_book} Following this approach, de~Haan and Slater 
have noted that the free energy function for a pore of vanishing length
is given by\cite{deHaan_2011a}
\begin{eqnarray}
\beta F(s/N) = (1-\lambda)\ln\left[ (1-s/N)(s/N)\right],
\label{eq:FsN}
\end{eqnarray}
where 
%
%
$s$ is the number of monomers that have 
translocated through the pore. Thus, the free energy is a universal function 
of the scaled variable $s/N$.
In the case where the diffusion coefficient $D$ in the FP equation 
is independent of $s$ (i.e. pore friction is dominant), the FP formalism
predicts\cite{deHaan_2011a} that such a free energy corresponds to a translocation first 
mean passage time that scales as $\langle\tau\rangle\sim N^2$. 

For a finite-length nanopore, the situation is somewhat more complicated, since
$D$ (and thus $\gamma_{\rm pol}$) changes during the pore-filling 
and emptying stages. Thus, $D$ is constant only during the stage when the pore is 
filled, where $s\in[0,N']$, where $N'\equiv N-1-n_{\rm p}$ is equal to the total
number of bonds that lie outside the nanopore. In this domain, it is straightforward 
to show that the free energy satisfies a form similar to that of Eq.~(\ref{eq:FsN}): 
\begin{eqnarray}
\beta F(s/N') = (1-\lambda)\ln\left[ (1-s/N')(s/N')\right],
\label{eq:FsN2}
\end{eqnarray}
neglecting terms constant with respect to $s$. Thus, during the filled-pore stage, 
the free energy is predicted to be a universal function of $s/N'$.
Note that this result can be shown to be consistent with the second term in
Eq.~(40) of Ref.~\onlinecite{Polson_2013}.
The polymer first reaches either bound in this range at the time $\tau_1$, as defined
earlier. The quantity $N'$ can be considered the effective length
of the polymer for the filled-pore stage of translocation. 
%
%

\begin{figure}[!ht]
\includegraphics[width=0.40\textwidth]{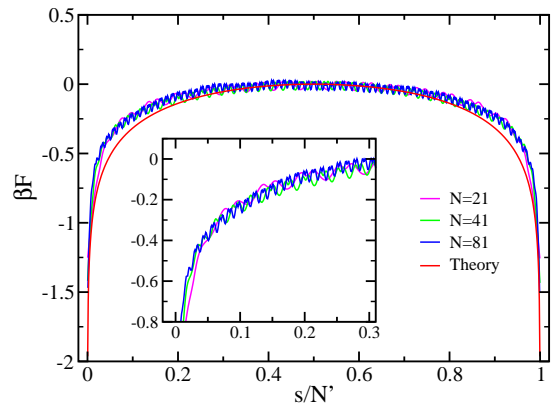}
\caption{Free energy vs $s/N'$. $s$ is the number of monomers
that have entered the pore just after the pore is filled, and 
$N'\equiv N-1-n_{\rm p}$ is
the number of bonds outside the nanopore, as described in the text.
The graph shows results for three different polymer lengths calculated
from MC SCMH simulations. Also shown is the theoretical prediction of
Eq.~(\ref{eq:FsN2}). The inset shows a close up near the edge of the
profile.}
\label{fig:F.N.scale}
\end{figure}

Figure~\ref{fig:F.N.scale} shows free energy functions for three different
polymer lengths plotted as a function of $s/N'$ for a nanopore of length $L=3$
and radius $R=0.65$. Note that functions are only shown for the filled-pore stage. 
In addition the curves have been shifted along the vertical axis so that the 
functions have approximately the same value at the mid-point of $s/N'=0.5$.
Also shown in the figure is the analytical prediction of Eq.~(\ref{eq:FsN2}).
The simulation curves each display weak oscillations that were noted earlier.
Note that the pore length $L$ has been chosen to minimize the oscillation amplitude.
The oscillations are more clearly visible in the enlargement in the inset of the figure.
Since the period of the oscillations is $\Delta s =1$, the oscillation period with 
respect to the scaled coordinate,  $s/N'$, decreases as $N$ increases. The model 
used to provide the analytical prediction does not incorporate the feature that 
gives rise to the oscillations, and so the theoretical curve is smooth, as expected.
Apart from the oscillations, the calculated free energy functions overlap
to a remarkable degree for the range of $N$ considered.  The theoretical curve
is close to, but somewhat narrower than the simulation curves.  This slight
discrepancy is probably due to a finite size effect, since the derivations of
Eqs.~(\ref{eq:FsN}) and (\ref{eq:FsN2}) use Eq.~(\ref{eq:Z}), a scaling relation
is valid in the limit of very long polymers. It is expected that the calculated
free energies will gradually approach the theoretical prediction with as $N$
is increased to much larger values.

The calculated free energy functions can be used to solve the FP equation
in order to calculate translocation time distributions. These calculations
also use the expressions for the friction coefficient, $\gamma_{\rm pol}$, derived
in Section~\ref{sec:theory}. However, this approach is valid only when 
the pore friction dominates and when the polymer is in conformational equilibrium.
One useful means to determine where these conditions hold is to examine the
scaling of the translocation time with chain length. Simulation studies of
translocation generally yield power laws of the form $\langle\tau\rangle\sim N^\alpha$.
Measurements of the exponent $\alpha$ have yielded a range of values, and the exponent
appears to be sensitive to the pore geometry and finite-size effects. When the translocation
time is sufficiently long relative to the relaxation time of the chain, the
value predicted using the FP method is expected. For unforced translocation,
the FP method gives $\alpha=2$, while in the case of a strong driving
force, $\alpha=1$. Recently, simulations of unforced translocation observed
$\alpha=2$, in one case by decreasing the viscosity of the solvent (and thus
shortening the polymer relaxation time),\cite{deHaan_2012a} and in the other
case by increasing the pore friction.\cite{Polson_2013b} In the present model,
a sufficiently large pore friction, $\gamma_{0{\rm p}}$, should slow translocation
to the point where the FP result is observed.

As noted earlier, the FP prediction of $\alpha=2$ for unforced translocation
is expected for two conditions: (i) a constant $D$ (i.e. constant $\gamma_{\rm pol}$)
in the FP equation, and (ii) the free energy is a universal function of either $s/N$
(pores of zero length) or $s/N'$ (finite-length pores).
For a finite-length pore, $D$ is constant in the filled-pore regime, within which
$F$ is approximately a universal function of $s/N'$ for the range of $N$ considered
in Fig.~\ref{fig:F.N.scale} ($21\leq N\leq 81$), though slightly different from the 
form predicted by Eq.~(\ref{eq:FsN2}).  The oscillations in the curves for $L=3$ 
in Fig.~\ref{fig:F.N.scale} are probably small enough to have a negligible effect 
on the translocation dynamics.  Thus, the appropriate scaling test for quasi-static 
conditions for finite-length pores is
\begin{eqnarray}
\langle\tau_1\rangle \propto (N-N_{\rm p})^\alpha,
\label{eq:tau1scale}
\end{eqnarray}
where the time $\tau_1$ is the first time where the monomers have completely emptied 
from the {\it cis} or {\it trans} domains (as illustrated in Fig.~\ref{fig:spic}(b) and (c)).  
Over this time interval, the pore remains 
filled and $D$ is constant. Also, $N_{\rm p}$ is the average number of monomers that 
lie inside a filled nanopore.  For $R=0.65$ and $L=3$, $N_{\rm p}=4.125$. 
Thus, the quantity $N-N_{\rm p}$ is the effective length of the polymer during
this stage of translocation (i.e. number of monomers that pass through the nanopore
during this stage of translocation).  In the regime of
sufficiently long chains and/or short pore length, the exponents obtained from
this relation should be identical to those from $\langle\tau\rangle\propto N^\alpha$.
%
%

\begin{figure}[!ht]
\includegraphics[width=0.40\textwidth]{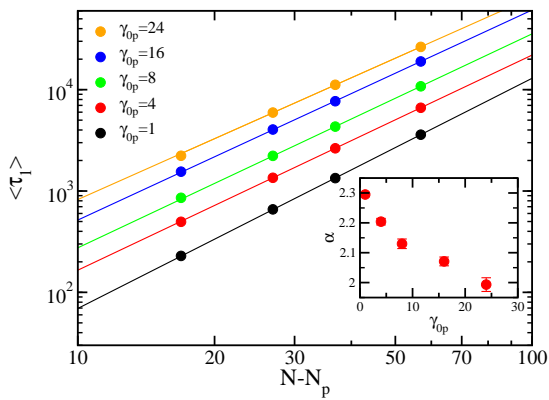}
\caption{Mean translocation time, $\langle\tau_1\rangle$, vs $N-N_{\rm p}$,
for several different values of pore monomer friction coefficient, $\gamma_{0{\rm p}}$.
The polymer starts midway in the pore, with $s_0=(N-1-n_{\rm p})/2$. In addition,
$L=3$, $R=0.65$, and $f_{\rm d}=0$. The inset shows the translocation time scaling 
exponent, $\alpha$, vs $\gamma_{0{\rm p}}$.}
\label{fig:tau.N.L3.R0.65.Q0.5.F0}
\end{figure}

Figure~\ref{fig:tau.N.L3.R0.65.Q0.5.F0} shows the scaling of $\langle\tau_1\rangle$
with chain length for several different values of the pore friction parameter, 
$\gamma_{0{\rm p}}$. As expected, decreasing $\gamma_{0{\rm p}}$ increases the 
translocation time.  Over the range of chain lengths considered ($N=21$--$61$),
the power law of Eq.~(\ref{eq:tau1scale}) is satisfied. In addition, as shown
in the inset of the figure, $\alpha$ varies with $\gamma_{0{\rm p}}$.
At $\gamma_{0{\rm p}}=\gamma_0=1$, $\alpha=2.30$. As $\gamma_{0{\rm p}}$ 
increases and the pore friction becomes increasingly dominant, $\alpha$ monotonically 
decreases.  At $\gamma_{0{\rm p}}=24$, the exponent is $\alpha=1.99\pm 0.02$, 
which is consistent with the FP prediction and indicates that the system is in
the quasi-static regime.  At higher $\gamma_{0{\rm p}}$, the system will remain
in this regime, and the exponent is expected to maintain a value close to $\alpha=2$. 
This scaling behaviour is generally consistent with that of the MC dynamics simulations 
in Ref.~\onlinecite{Polson_2013b}, where the pore friction was tuned by varying the 
pore radius. In addition, these results closely parallel those of 
Ref.~\onlinecite{deHaan_2012b}, where decreasing the solvent viscosity
in that study has the same effect on $\alpha$ as increasing the pore friction
in the present work.  It is important to note that the measured exponents
are valid only for the range of polymer lengths considered here. 
As the polymer length increases, the relaxation time of the polymer also increases, 
and a greater $\gamma_{0{\rm p}}$ will be required to maintain quasi-static
conditions. For any pore friction corresponding to this regime for short polymers,
increasing $N$ will eventually push the system out of equilibrium, and
the scaling exponent will change. 

\begin{figure}[!ht]
\includegraphics[width=0.40\textwidth]{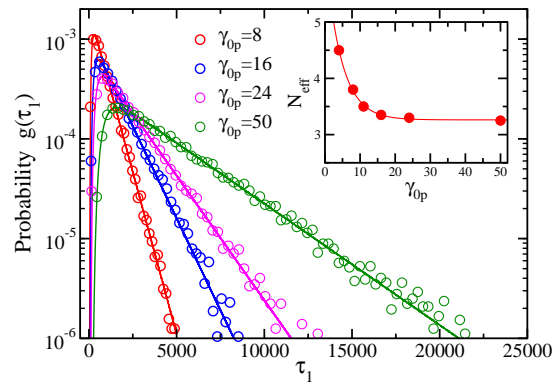}
\caption{Translocation time distributions for unforced translocation where
the polymer starts midway in the nanopore, with $s_0=(N-1-n_{\rm p})/2$.
In addition, $L=3$, $R=0.65$, and $f_{\rm d}=0$. The open circles denote BD simulation
data, and the solid curves show FP theoretical predictions. The inset shows
$N_{\rm eff}$ vs $\gamma_{0{\rm p}}$, for $N_{\rm eff}$ values that provide the
best match between the FP predictions and BD data. The solid line is a guide for
the eye.}
\label{fig:taudist.N21.R0.65.L3.Q0.5.F0.type1}
\end{figure}

Translocation time distributions calculated from BD simulations for a $N=21$ polymer 
are shown in Fig.~\ref{fig:taudist.N21.R0.65.L3.Q0.5.F0.type1}. The distributions stretch
to higher translocation times as the pore friction increases, as expected.
Overlaid on the BD curves are distributions predicted using the FP method.
As explained in Section~\ref{sec:theory}, the friction coefficient, $\gamma_{\rm pol}$,
is parameterized by $N_{\rm eff}$. This quantity is the effective number of monomers
whose dynamics is governed by friction inside the pore. The theoretical curves
have been calculated using values of $N_{\rm eff}$ that provide the best match
between the theoretical and BD distributions. The theoretical curves are in excellent 
agreement with the simulation data.  For lower values of the pore friction,
the scaling results of Fig.~\ref{fig:tau.N.L3.R0.65.Q0.5.F0} yield $\alpha > 2$,
indicating that at least one of the two conditions for FP validity is not met.
This fact is also evident in the variation of $N_{\rm eff}$ with pore friction
shown in the inset of the figure. $N_{\rm eff}$ decreases with $\gamma_{0{\rm p}}$
but levels off to a value near $N_{\rm eff}\approx 3.25$ at high $\gamma_{0{\rm p}}$.
This value is close to the approximation in Eq.~(\ref{eq:NeffLas}) and
even closer to the value of $N_{\rm eff}=3.2$ measured directly from MC simulations, 
as described in Section~\ref{sec:theory}.  As noted earlier, any appreciable deviation from this
prediction indicates a breakdown of the condition of conformational quasi-equilibrium 
and/or the dominance of pore friction. As the pore friction increases in strength,
these two conditions become increasingly valid, as evident from the $N_{\rm eff}$
values. These results and conclusions are consistent with those observed in
our recent MC dynamics translocation study.\cite{Polson_2013}

It is a curious fact that it is possible for the FP distributions to be 
in such excellent agreement with the BD distributions even outside the regime where 
the conditions required for the theory to be valid hold (e.g. at $\gamma_{0\rm p}=8$).  
Thus, the functional form of the measured distributions provide no clear indication 
of where the correct regime lies. This demonstrates the importance of employing 
complementary measurements of other relevant quantities, such as $\alpha$. It also 
demonstrates the importance of establishing a clear relation between the friction 
coefficient used in the FP equation and the related parameter(s) employed in the 
simulation model.

\begin{figure}[!ht]
\includegraphics[width=0.40\textwidth]{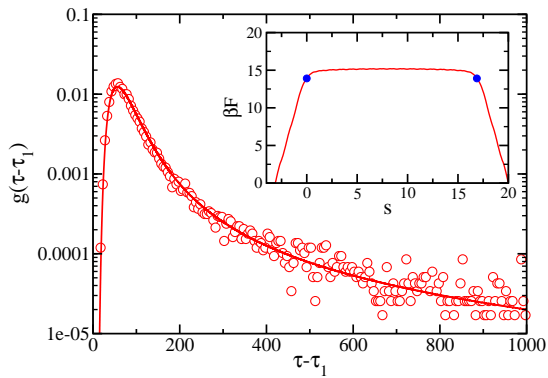}
\caption{Distribution of the pore emptying time $\tau-\tau_1$ for unforced translocation.
In this stage, the polymer starts at $s=0$ at time $\tau_1$ and proceeds until the time 
$\tau$, at which the nanopore is fully empty.  The monomer pore friction is $\gamma_{0\rm p}=50$.
In addition, $N=21$, $L=3$ and $R=0.65$. The open circles denote BD simulation
data, and the solid curve shows the FP theoretical prediction. The inset shows
the free energy curve used in the calculation. The blue circles indicates the two 
possible starting points of the pore emptying stage.  }
\label{fig:taudiffdist.N21.R0.65.L3.Q0.5.F0.G50}
\end{figure}

Now consider the pore emptying stage for unforced translocation. At time $t=\tau_1$,
the polymer has just reached the point where the {\it cis} or {\it trans} domain
has emptied, while at time $t=\tau$, the nanopore itself has just emptied. Thus,
the pore emptying time can be defined as difference $\tau-\tau_1$. 
Figure~\ref{fig:taudiffdist.N21.R0.65.L3.Q0.5.F0.G50} shows this time distribution 
for an $N=21$ polymer moving through a pore with friction $\gamma_{0\rm p}=50$.
From the result in Fig.~\ref{fig:taudist.N21.R0.65.L3.Q0.5.F0.type1}, it is clear that
this corresponds to a case where the polymer maintains conformational quasi-equilibrium during the
filled-pore stage. However, it is not clear that this condition will be maintained
during pore emptying. As is evident in the inset of the figure, which shows the 
free energy profile for the system, the free energy gradient is much larger here
than during the filled stage. Consequently, the translocation rate is expected
to be much greater, and this could result in conformational distortion of
the {\it cis} and {\it trans} subchains. To test this condition, FP calculations
were carried out using $s=0$ as the initial state. The polymer friction 
$\gamma_{\rm pol}(s)$ for the emptying stage was calculated using the method 
described in Section~\ref{sec:theory}. The calculations used the same value
of $N_{\rm eff}=3.25$ used in the fit to the $\gamma_{0\rm p}=50$ curve in 
Fig.~\ref{fig:taudist.N21.R0.65.L3.Q0.5.F0.type1}.  The calculated distribution is 
shown as the solid curve in the figure, and is in excellent agreement with the results
of the simulation. Evidently, the friction is large enough in this case to maintain
quasi-static conditions.

\begin{figure}[!ht]
\includegraphics[width=0.40\textwidth]{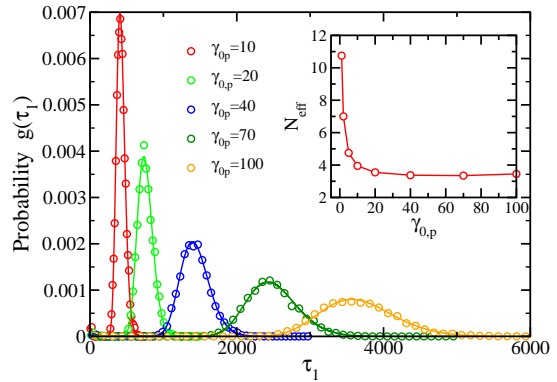}
\caption{Translocation time distributions for driven translocation for several values of 
$\gamma_{0{\rm p}}$.  The symbols denote BD simulation data  
and the solid curves show theoretical predictions using the FP formalism.  The data are 
for a polymer of length $N=41$ starting at $s_0=2.04$ and subject to a driving force 
of $f_{\rm d}=1$. In addition, the pore dimensions are $L=3$ and $R=0.65$.
The inset shows $N_{\rm eff}$ vs $\gamma_{0{\rm p}}$. These values of $N_{\rm eff}$
provided the best match between the FP and BD distributions.}
\label{fig:taudist.N41.R0.65.L3.Q0.1.F1.type1}
\end{figure}

Next, we consider the case of forced translocation. 
Figure~\ref{fig:taudist.N41.R0.65.L3.Q0.1.F1.type1}
shows translocation time distributions for a $N=41$ polymer subject to a driving
force of $f_{\rm d}=1$. Results for several different different values of $\gamma_{0{\rm p}}$
are shown. In these simulations, the polymer was initially placed at $s_0=2.04$.
As evident in Fig.~\ref{fig:F.N41.L3.R0.65}(b), this position is on the {\it trans}
side of the free energy maximum located near $s=-0.33$. The translocation was sufficiently 
advanced to prevent the polymer exiting on the {\it cis} side.  Most of the monomers 
initially lie in the {\it cis} compartment and are pushed through the nanopore during 
translocation.

The distributions are Gaussian-like, and they shift to longer times
as the pore friction increases, as expected. In addition, the distribution width
also increases with the pore friction strength. Overlaid on the BD distributions are 
theoretical curves calculated using the FP approach. The $N_{\rm eff}$ values used
in the FP calculations are shown in the inset of the figure.
As in the case of unforced translocation, the theoretical curves are in excellent
agreement with the BD distributions for the data shown. In this case, 
quantitative agreement was not especially good at low pore friction, 
$\gamma_{0{\rm p}}<10$ (data not shown). These results are easily understood 
from examining the data in the inset. Consistent with the results shown in the inset of
Fig.~\ref{fig:taudist.N21.R0.65.L3.Q0.5.F0.type1}, $N_{\rm eff}$ is very high
at low $\gamma_{0{\rm p}}$, but decreases rapidly and levels off at a value of
$N_{\rm eff}=3.35$ at high $\gamma_{0{\rm p}}$. This limiting value is close to 
the expected value of $N_{\rm eff}=3.2$. The slight higher value measured
from the simulation data is probably due to a residual contribution to the total
polymer friction from the monomers outside the pore. To summarize, the FP 
distributions are in excellent agreement with the simulation data in the high 
pore-friction regime, where the theory is expected to work, and in poorer 
agreement with simulation outside this regime. 

It is instructive to examine the conformational behaviour of the polymer during 
translocation directly. One useful measure is the displacement of the
end monomer on the {\it cis} and {\it trans} sides from the centre of the nearest
opening of the nanopore. The $z$ components of these displacements are
$Z_{\rm c} = z_1$ and $Z_{\rm t} = z_N - L$ respectively,
where $z_1$ is the $z$ coordinate of monomer 1, defined to be the last to go through
the nanopore, and $z_N$ is the coordinate of monomer $N$, defined to be the first
to go through the pore. 

\begin{figure}[!ht]
\includegraphics[width=0.40\textwidth]{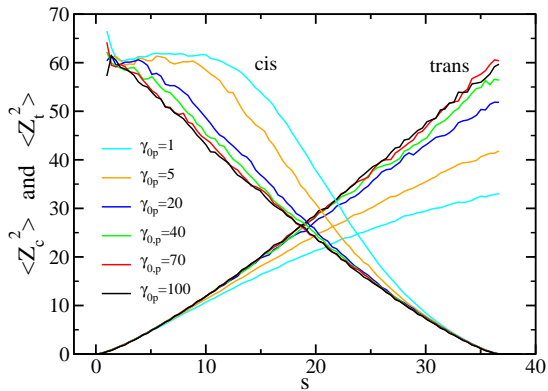}
\caption{$\langle Z_{\rm c}^2\rangle$ and $\langle Z_{\rm t}^2\rangle$ of the 
{\it cis} and {\it trans} subchains as a function of $s$ several different 
values of $\gamma_{0{\rm p}}$.  The data are averages over many translocation events 
for a polymer of length $N=41$ starting at $s_0=2.04$ and subject to a pore monomer 
driving force of $f_{\rm d}=1$.  In addition, $L=3$ and $R=0.65$.}
\label{fig:rez_ct.N41.R0.65.L3.Q0.1.F1}
\end{figure}

Figure~\ref{fig:rez_ct.N41.R0.65.L3.Q0.1.F1} shows variation of $\langle Z_{\rm c}^2\rangle$ 
and $\langle Z_{\rm t}^2\rangle$ with $s$ during translocation. 
These results were obtained from the simulations used in
Fig.~\ref{fig:taudist.N41.R0.65.L3.Q0.1.F1.type1}.
Results for several different $\gamma_{0{\rm p}}$ are shown.
At low $\gamma_{0{\rm p}}$, where translocation is rapid, there is a compression
of the {\it trans} subchain. On this {\it cis} side, there is a corresponding
lag in the response of the position of the polymer end to the pull of the
polymer during translocation.  These results are a direct measure of the
out-of-equilibrium character of the polymer. The delayed response on the
{\it cis} side is in keeping with the finite time required for the ``tension front''
to reach the {\it cis } end in the context of the tension propagation theory
of translocation.\cite{Sakaue_2007,Sakaue_2010,Rowghanian_2011,Dubbeldam_2012,%
Ikonen_2012a,Ikonen_2012b} As the pore friction increases, 
$\langle Z_{\rm t}^2\rangle$ monotonically increases and $\langle Z_{\rm c}^2\rangle$ 
decreases. Thus, the polymer tends toward conformational equilibrium. At 
$\gamma_{0{\rm p}}\approx 70$, the polymer has reached conformational quasi-equilibrium. 
This is evident from the observed symmetry between the {\it cis} and {\it trans}
curves at high $\gamma_{0{\rm p}}$. Results of comparable calculations using
the $x$ and $y$ components of the end-monomer displacements, as well as the
radius of gyration of the {\it cis} and {\it trans} subchains, showed identical
trends (data not shown). Finally, it should be noted that increases in both the polymer
length and the driving force are expected push the system out of equilibrium.
Thus, the degree of pore friction required to achieve quasi-equilibrium will 
increase as $N$ increases and as $f_{\rm d}$ increases.

A close comparison of the results of Figs.~\ref{fig:taudist.N41.R0.65.L3.Q0.1.F1.type1}
and \ref{fig:rez_ct.N41.R0.65.L3.Q0.1.F1} reveals an interesting point.
Both sets of data show that conformational quasi-equilibrium is approached as
the pore friction strength increases. However, the variation of $N_{\rm eff}$ with
$\gamma_{0{\rm p}}$ shows that the FP formalism appears to be valid at
$\gamma_{0{\rm p}} = 40$ and higher, while the direct measurement of the conformational
state of the polymer shows the polymer is still somewhat out of equilibrium at 
$\gamma_{0{\rm p}}=40$. This shows that agreement between the FP predictions and simulation data 
(and, by extension, experimental data) does not provide a perfectly accurate test
for the quasi-equilibrium condition as the system crosses over into this regime.

\begin{figure}[!ht]
\includegraphics[width=0.40\textwidth]{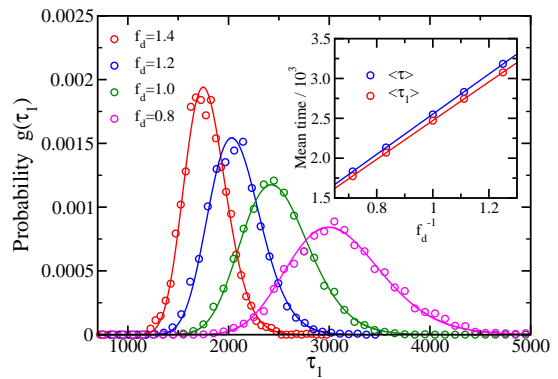}
\caption{Translocation time distributions for driven translocation for several values of
the driving force, $f_{\rm d}$.  The symbols denote BD simulation data and the solid curves 
show theoretical predictions. The data are for a polymer of length $N=41$ starting at $s_0=2.04$ 
and a pore with a friction of $\gamma_{0{\rm p}}=70$ and dimensions of $L=3$ and $R=0.65$.
The FP curves were calculated using $N_{\rm eff}=3.35$. The inset shows $\langle\tau_1\rangle$ 
and $\langle\tau\rangle$ vs $f_{\rm d}^{-1}$. The solid lines are fits to the data using 
a power law functional form. The fits yielded a scaling of $f_{\rm d}^{-0.984\pm 0.006}$ 
for both $\langle\tau_1\rangle$ and $\langle\tau\rangle$.  }
\label{fig:taudist.N41.R0.65.L3.Q0.1.G70.type1}
\end{figure}

Figure~\ref{fig:taudist.N41.R0.65.L3.Q0.1.G70.type1} shows translocation time distributions
for a $N=41$ polymer for various driving forces. The pore friction strength is 
$\gamma_{0{\rm p}}=70$. The variation of $\langle Z_{\rm c}^2\rangle$
and $\langle Z_{\rm t}^2\rangle$ with $s$ is the same for all values
of $f_{\rm d}$ considered here (data not shown), indicating that the polymer
is in conformational equilibrium over this range. Theoretical distributions
were calculated for $N_{\rm eff}=3.35$ and are all in excellent agreement
with the simulation data. The inset shows the variation of $\langle\tau_1\rangle$ 
and $\langle\tau\rangle$ with $f_{\rm d}^{-1}$. Fits to a power law  
$\langle\tau_1\rangle\propto f_{\rm d}^{\delta}$ yielded an exponent of 
$\delta = -0.984\pm 0.006$.  A fit of $\langle\tau\rangle$ to the same relation 
yielded the same scaling exponent.  These results are very close to the 
prediction that $\langle\tau\rangle \propto f_{\rm d}^{-1}$ by Muthukumar in 
the limit where pore friction dominates.\cite{Muthukumar_1999} 
Note once again that increases in the both the driving force and the polymer length
are expected to push the system out of equilibrium, leading to dynamical
behaviour that diverges from that of the FP predictions. This is illustrated
by recent MD simulations for a $N=128$ polymer, which yield values of 
$\delta\approx -0.9$ for low and intermediate forces, and somewhat higher values 
(that depend on the details of the model) for strong forces.\cite{Ikonen_2012b}

Finally, we examine the scaling of the translocation time with polymer length, $N$.
In the limit of very strong driving force and zero pore length, the FP approach
predicts\cite{Muthukumar_1999,Muthukumar_book} a scaling of $\langle\tau\rangle\propto N$. 
This corresponds to the case where the free energy function is dominated by the chemical 
potential difference across the pore and where the entropic barrier contribution is negligible.
In this limit, the mean velocity is constant and thus the mean coordinate varies as 
$\langle s\rangle \propto t$, where $t$ is the time measured from the beginning of 
the translocation process. Equivalently, it is expected that $\langle s-s_0\rangle \propto t$ 
for an arbitrary initial coordinate $s_0$. For translocation of chains of arbitrary length
each located at the same $s_0$ at $t=0$, it is then predicted that 
$\langle\tau_1\rangle\propto\Delta s$, where $\Delta s = N-1-n_{\rm p}-s_0$
is the displacement during the time interval $[0,\tau_1]$

\begin{figure}[!ht]
\includegraphics[width=0.40\textwidth]{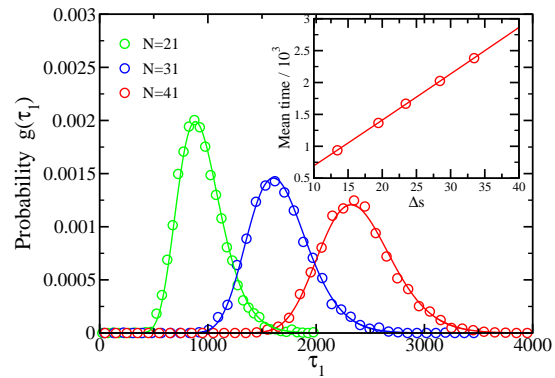}
\caption{Translocation time distributions $g(\tau_1)$ for driven translocation 
for polymers of several lengths. In each case translocation starts at $s_0=3.44$.
In addition, $\gamma_{0\rm p}=70$ and $f_{\rm d}=1$. The symbols denote BD 
simulation data and the solid curves show theoretical predictions. 
The FP curves were calculated using $N_{\rm eff}=3.35$. The inset shows 
$\langle\tau_1\rangle$ vs $\Delta s$, where $\Delta s=N-1-n_{\rm p}-s_0$.
The solid line is a fit to the data using a power law. The fit yielded
a scaling of $\langle\tau_1\rangle \propto \Delta s^{1.025\pm 0.004}$. }
\label{fig:taudist.ifix5}
\end{figure}

To test this prediction, we measured translocation times $\tau_1$ for chains of various lengths
for $\gamma_{0\rm p}=70$, $f_{\rm d}=1$, and an initial coordinate value of $\Delta s=3.44$.
For these conditions and the range of lengths considered ($N=21$--$41$) we know from the results 
shown in the figures above that the polymer will remain in conformational quasi-equilibrium.
The inset of Fig.~\ref{fig:taudist.ifix5} shows that $\langle\tau_1\rangle$ does appear 
to vary linearly with $\Delta s$. A power law fit yields a scaling of 
$\langle\tau_1\rangle\propto\Delta s^{1.025\pm 0.004}$. The slight deviation of
the scaling exponent from unity may be a residual effect of the entropic contribution
to the free energy barrier.  Figure~\ref{fig:taudist.ifix5} also shows translocation time 
distributions for the different chain lengths. Overlaid are the theoretical predictions using 
the FP approach and the same  value of $N_{\rm eff}=3.35$ used for calculations with
results shown in Fig.~\ref{fig:taudist.N41.R0.65.L3.Q0.1.G70.type1}. The predictions
are in excellent agreement with the simulation results. 

The FP scaling exponent for driven translocation of $\alpha=1$ is only valid in
a regime that is determined by values of the pore friction, polymer length and 
driving force. Outside the quasi-equilibrium regime, $\alpha$ is expected to deviate 
from this value.  As noted earlier, BDTP calculations have shown that $\alpha$ varies 
with each these parameters, approaching a value of $\nu = 1+\nu\approx 1.6$ in the 
limit of large $N$.\cite{Ikonen_2012b} In the limit of high pore friction and 
low $N$, $\alpha$ tends toward a value of unity, suggesting that the FP regime 
considered in this study may be a limiting case of that theoretical model.

\section{Conclusions}
\label{sec:conclusions}

In this study, we have used BD simulations to study the translocation dynamics of a
coarse-grained polymer. These simulations differ from those of most other studies
by the presence of a tunable friction parameter, $\gamma_{0{\rm p}}$, for monomers
that lie inside the pore. We have considered the case where $\gamma_{0{\rm p}} \geq \gamma_0$,
where $\gamma_0$ is the friction outside the pore, whereas other studies have essentially
used $\gamma_{0\rm p}=\gamma_0$. Increasing $\gamma_{0\rm p}$ reduces the rate of translocation. 
For low $\gamma_{0\rm p}$, where translocation is rapid, the polymer conformation exhibits
out-of-equilibrium conformational behaviour, while for sufficiently high $\gamma_{0\rm p}$ 
and slow translocation, a state of quasi-equilibrium is observed. We have examined the
behaviour of the translocation time $\tau_1$ on the pore friction, and have analyzed
the data using the FP formalism together with translocation free energy functions that
have been calculated explicitly using MC simulations.

Generally, the analysis of the data yield results that are consistent with the predictions
of the FP approach for sufficiently strong pore friction, and poorer results for 
weaker pore friction.  In the case of unforced translocation, the scaling exponent
for $\langle\tau_1\rangle$ was measured to be $\alpha=1.99\pm 0.02$ for strong friction.
This result has also been observed in Refs.~\onlinecite{deHaan_2012a} and 
\onlinecite{Polson_2013b} and is expected in the case where pore friction dominates 
and conformational quasi-equilibrium holds.
In addition, for translocation driven by a strong force of strength $f_{\rm d}$, we observed
scalings very close to $\langle\tau_1\rangle\propto f_{\rm d}^{-1}$ at sufficiently high $\gamma_{0\rm p}$.
Upon variation in the chain length $N$, we observed a scaling of approximately 
$\langle\tau_1\rangle\propto \Delta s$, where $\Delta s$ is the translocation coordinate range 
between the initial value and the value when the {\it cis} region has just emptied.
This provides indirect evidence for the validity of the scaling predicted by the FP approach of
$\langle\tau\rangle\propto N$ in the limit of strong force and strong pore friction.
As another test, FP calculations were used to analyze the translocation time distributions 
obtained from the BD simulations. These calculations use a total pore friction value that 
is parameterized by a quantity $N_{\rm eff}$, the effective number monomers whose dynamics
are strongly affected by the pore. For sufficiently strong pore friction, the FP distributions 
were in excellent agreement with those obtained from the simulations using physically 
meaningful values of $N_{\rm eff}$.

The results of this study provide direct evidence that the FP formalism provides a valid
description of polymer translocation dynamics in the regime where the translocation time 
is sufficiently long relative to the conformational relaxation time of the polymer. 
The pore friction strength required to maintain quasi-equilibrium is expected to increase
with increasing polymer length and driving force. It will also be affected by
incorporation of hydrodynamic correlations, which will change the relaxation time
of the polymer.  Delineation of the regime of validity 
for the FP equation in describing translocation dynamics would be illuminating but time 
consuming and is beyond the scope of the present study.

\begin{acknowledgments}
JMP would like to thank Sheldon B. Opps for helpful discussions.
This work was supported by the Natural Sciences and Engineering Research Council of Canada (NSERC).
We are grateful to the Atlantic Computational Excellence Network (ACEnet) for use of 
their computational facilities.
\end{acknowledgments}

%
%


%

\end{document}